\documentclass[twocolumn,showpacs,preprintnumbers,amsmath,amssymb]{revtex4}

\usepackage{graphicx}
\usepackage{dcolumn}
\usepackage{bm}

\begin{document}

\title{Space Charge Influence on the Angle of Conical Spikes \\
       Developing on a Liquid-Metal Anode}

\author{G. Sh. Boltachev} \email{grey@iep.uran.ru}
\author{N. M. Zubarev}    \email{nick@ami.uran.ru}
\author{O. V. Zubareva}   \email{olga@ami.uran.ru}
\affiliation{Institute of Electrophysics, Ural Division, 
  Russian Academy of Sciences, \\ 
  106 Amundsen Street, 620016 Ekaterinburg, Russia}

\begin{abstract} 
The influence of the space charge of ions emitted from the surface of 
  a conical spike on its shape has been studied. 
The problem of the calculation of the spatial distributions of the electric field,
  ion velocity field, and the space charge density near the cone tip 
  has been reduced to the analysis of a system of ordinary differential equations. 
As a result of numerical solution of these equations, the criterion of the balance
  of the capillary and electrostatic forces on the conic surface of a liquid-metal
  anode has been determined.
It has allowed us to relate the electrical current flowing through the system, 
  the applied potential difference and the cone angle. 
We have compared the results of our calculations with available experimental data
  concerning emission from the surface of pure liquid gallium (Ga), indium (In),
  tin (Sn), and some liquid alloys, such as Au$+$Si, Co$+$Ge, and Au$+$Ge. 
On the basis of the proposed model, explanations have been given for a number 
  of specific features of the emissive behavior of different systems. 
\end{abstract}


\maketitle

\section*{Introduction}

It is known that, as a result of the electrohydrodynamic instability development,
  the surface of a conducting liquid (liquid metal) takes the shape of a cone
  in a rather strong external electric field 
  \cite{l.Zel,l.Taylor,l.Forbes,l.Odder,l.Zub01,l.Zub001,l.Zub04}. 
An enhancement of the field near the cone tip provides conditions for 
  the initiation of emission processes such as field evaporation of ions 
  \cite{l.Gor,l.Kan,l.Kin,l.Cho,l.Wang}. 
Interest in studying the geometry of such structures was stimulated in large 
  measure by the development of liquid-metal ion sources (LMIS).
Considerable progress in the theory of conical spikes started from Taylor's 
  investigations \cite{l.Taylor,l.Tay2}. 
He has shown that, for a cone with a half-angle of $\alpha_T\simeq 49.3^{\circ}$, 
  the surface electrostatic pressure $P_E$ depends on the distance from its apex 
  as $R^{-1}$ and, hence, can be counterbalanced by the surface pressure $P_L\sim R^{-1}$.

The geometry of ion-emitting cones (Taylor cones) was investigated in numerous 
  experimental works \cite{l.In,l.Ga,l.AuSi,l.Sn,l.CoGe,l.AuGe}. 
These works testify that an increase in the applied potential difference is 
  accompanied not only by the appearance and increase of the emission current,
  but also by a decrease of the cone half-angle.
For small currents, i.e., when the space charge influence is negligible, 
  the half-angle is close to Taylor's angle $\alpha_T$. 
The phenomenon of a cone sharpening can be interpreted as the system response
  aimed at the conservation of the balance of the pressures $P_E$ and $P_L$ under
  the conditions of the screening effect of the space charge. 
The last effect in case of an emitter with invariable shape, 
  as is known \cite{l.Child,l.Langmuir}, reduces to the Child-Langmuir law.

Simple analytic models, which relate the basic parameters of a problem, play 
  an important role in gaining an insight into the physical processes 
  that occur in liquid-metal ion sources. 
Among these models are Mair's theory \cite{l.Mair1}, the models by Kingham 
  \& Swanson \cite{l.Kin}, Mair \& Forbes \cite{l.Mair2,l.Mair3}, and, certainly,
  Taylor's model \cite{l.Taylor}.
Nevertheless, the above-listed models do not present a theoretical description
  of the current dependence of the cone angle. 
In this work, we propose a model, which generalizes, on the one hand, Taylor's 
  solution to the case where the space charge starts playing an important part,
  and on the other hand, the Child-Langmuir law to the case of an emitter 
  of variable (self-adjusting) shape. 
It is based on self-similar solutions for a charged particle flow \cite{l.Kir,l.Finn},
  which were found to be compatible with the Laplace-Young stress condition 
  for a liquid conducting cone: see our Letter \cite{l.BZ06}. 
Clearly, this model does not attempt to describe all the features of operation
  of liquid-metal ion sources. 
The application of self-similar solutions restricts our analysis to precisely 
  conical shape of the emitter. 
So, possible deformations of the emitting cone observed in experiments, in particular, 
  the appearance of small jet-like protrusion on its vertex remain beyond the scope
  of this paper. 
The advantage of the proposed model is the possibility to describe distributions
  of the electric field potential and of the ion velocity field over the cone and,
  as a consequence, to obtain relations between the cone angle, the current,
  and the applied voltage.

\section{Initial equations; self-similar reduction}

\begin{figure}
\centering
\includegraphics*{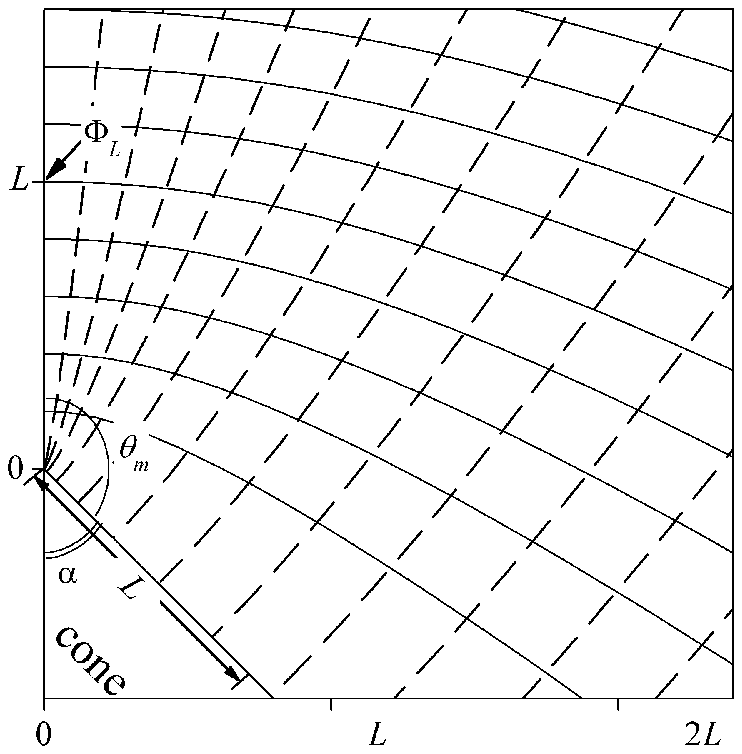}
\caption{The geometry of the problem and the notations: 
  $L$ is the characteristic spatial scale,
  $\alpha$ is the cone half-angle, $\theta_m \equiv \pi - \alpha$, 
  $\Phi_L$ is the absolute value of electric field potential at the symmetry axis
  at the distance $L$ from the apex of the cone.
  The solid lines correspond to the equipotential surfaces, including the surface
  of the cone with $\Phi = 0$. 
  The dashed lines correspond to the streamlines of the ion flow.}
\label{f.G01}
\end{figure}

Let us consider a single-velocity flow of ions evaporating from the surface
  of an infinite conical anode in the framework of the hydrodynamic description
  \cite{l.Kir,l.Finn,l.BZ06,l.LLGD}. 
Figure \ref{f.G01} shows the geometry of the problem and the notations used.
The mass and charge densities are proportional to each other for 
  a charged-particle flow, and vorticity of the flow is conserved 
  in a potential electric field (${\bf E}=-\nabla\Phi$). 
We assume the initial ion velocity to be equal to zero.
This approximation can be considered to be well founded in view of 
  the extremely high values of electric field strength near the cone tip. 
Then the vorticity of the flow is zero close to the anode surface, and, 
  hence, the ion flow is potential. 
In this case, the flow velocity, along with the electric field, can be described
  with the help of a scalar function, the velocity potential: ${\bf v}=\nabla\Psi$. 
Then the original set of equations can be written as follows:
\begin{eqnarray}
&& \nabla^2\Phi = - q N/\varepsilon_0,         \label{system1} \\
&& m\left(\nabla\Psi\right)^2\! / 2 = - q\Phi, \label{system2} \\
&& \nabla \left(N\nabla\Psi\right) = 0.        \label{system3}
\end{eqnarray}
The first equation is the Poisson equation for the electric field potential $\Phi$;
  here $N$ is the charged particle density, $q$ is the particle charge, 
  and $\varepsilon_0$ is the vacuum permittivity. 
The motion of particles is described by the second equation, which gives the 
  energy conservation law for ions in an electric field ($\Psi$ is the velocity
  potential, $m$ is the mass of a particle).
The third equation is the continuity equation. 
The Poisson equation (\ref{system1}) has to be solved together with the condition
  that the conductor surface is equipotential. 
It is convenient to take the electric potential of the the anode be equal to zero,
  $\Phi = 0$, which agrees with the condition that the initial velocity of particles
  is zero, $\nabla\Psi = 0$ (see equation (\ref{system2})).

The equilibrium configuration of the free liquid-metal surface is determined by 
  the balance condition for the electrostatic and capillary forces
  (the Laplace-Young equation). 
For a conical surface it takes the form: 
\begin{equation}
P_L = \frac{\sigma}{R} \cot{\alpha} = 
  \frac{\varepsilon_0}{2}\, \left(\nabla\Phi\right)_{\Phi=0}^{2} = P_E,
\label{equiv}
\end{equation}
  where $\alpha$ is the cone half-angle, $\sigma$ is the surface tension,
  and $R$ is the distance from the cone apex.
In the case of a high mass flow rate, characteristic for the so-called cone-jet
  mode of electrosprays \cite{l.Ganan,l.Barr} or for sonic sprays \cite{l.Hira},
  the hydrodynamic term also contributes to the pressure balance condition. 
However, for the process of field evaporation of ions realized in 
  the experiments \cite{l.In,l.Ga,l.AuSi,l.Sn,l.CoGe,l.AuGe},
  this term appears to be 2--3 orders of magnitude less than the
  electrostatic ($P_E$) and capillary ($P_L$) terms \cite{l.Forbes},
  and we can neglect it.

In order to pass to dimensionless variables, we introduce some characteristic
  spatial scale $L$, which will be taken as the unit of length (see Fig. \ref{f.G01}). 
Since the equations (\ref{system1})--(\ref{system3}) have no their own
  characteristic spatial size (they are invariant with respect to scaling),
  the value of $L$ cannot be defined from the model. 
Below we will treat $L$ as a size of the top part of an infinite model cone;
  the emission current from this part of the cone will be identified with the
  experimental current from a liquid-metal ion source. 
As an external control parameter, we introduce the absolute value of electric
  field potential at the symmetry axis at the same distance $L$ from the apex
  of the cone, $\Phi_L>0$. 
Then it is convenient to define the dimensionless variables by the following way:
\begin{equation}
\begin{array}{c}
r = R/L, \qquad \phi = - \Phi / \Phi_L, \\[1.5ex]
n = N L^2 q / \left(\varepsilon_0\Phi_L\right), \ \ 
\psi = \Psi \left[ m / \left(2 q L^2 \Phi_L\right) \right]^{1/2},

\end{array}
\label{undim}
\end{equation}
  and the initial equations (\ref{system1})-(\ref{system3}) takes the form:
\begin{equation}
\nabla^2\phi = n, \quad \left(\nabla\psi\right)^2 = \phi, \quad
  \nabla \left(n\nabla\psi\right) = 0.
\label{sysd}
\end{equation}
Switching to spherical coordinates and taking into consideration 
  the axial symmetry of the problem, we obtain:
\begin{equation}
\begin{array}{c}
\displaystyle \phi_{rr} + \frac{2}{r} \phi_r + \frac{1}{r^2} \phi_{\theta\theta}
    + \frac{\cot{\theta}}{r^2} \phi_\theta = n, \\[1.5ex]
\displaystyle \psi_r^2 + \frac{1}{r^2} \psi_\theta^2 =  \phi, \\[1.5ex]
\displaystyle n_r \psi_r + \frac{n_\theta\psi_\theta}{r^2} + n  \left[ \psi_{rr} + 
    \frac{2}{r} \psi_r + \frac{\psi_{\theta\theta}}{r^2} + 
    \frac{\cot{\theta}}{r^2} \, \psi_\theta \right] = 0.
\end{array}
\label{sysrt}
\end{equation}
It can readily be seen that these equations are invariant under the scale transformations 
\begin{equation}
\begin{array}{c}
\phi\to\phi\cdot s^{\gamma}, \qquad n\to n\cdot s^{\gamma-2}, \\[1.5ex]
\psi\to\psi\cdot s^{\gamma/2+1},\qquad r\to r \cdot s, 
\end{array}
\end{equation}
  where $\gamma$ and $s$ are some constants. 
Consequently, the system (\ref{sysrt}) can be reduced by means of the 
  following self-similar ansatz (see also \cite{l.Kir,l.Finn}):
\begin{equation}
\phi = r^{\gamma} A(\theta), \quad n = r^{\gamma-2} B(\theta), \quad 
   \psi = r^{\gamma/2+1} C(\theta), 
\end{equation}
  where $A$, $B$, and $C$ are unknown functions of the polar angle $\theta$. 

In the dimensionless form, the pressure balance condition (\ref{equiv}) reads
\begin{equation}
\left(\nabla\phi\right)^2_{\phi=0} = \frac{\cot\alpha}{rV^2}, \ \ \ \ 
V \equiv \Phi_L \sqrt{\frac{\varepsilon_0}{2L\sigma}}.
\label{eqdim}
\end{equation}
The dimensionless group $V$ plays the role of an external control parameter 
  of the problem. 
One can see from this expression that the electric potential has to depend 
  on the distance $r$ as $\phi\sim r^{1/2}$ that uniquely determines the value 
  of the self-similarity parameter: $\gamma=1/2$ \cite{l.Taylor,l.Tay2}. 
Thus, we should apply the substitution
\begin{eqnarray}
&& \phi\left(r,\theta\right) = r^{1/2} A(\theta), \nonumber\\
&& n\left(r,\theta\right) = r^{-3/2} B(\theta), \label{rasdel} \\ 
&& \psi\left(r,\theta\right) = r^{5/4} C(\theta), \nonumber
\end{eqnarray}
  which provides separation of variables in (\ref{sysrt}). 
The substitution (\ref{rasdel}) enables us to reduce the partial differential
  equations (\ref{sysrt}) to the following set of second-order ordinary 
  differential equations for the angle distribution of $A(\theta)$, $B(\theta)$
  and $C(\theta)$:
\begin{equation}
\begin{array}{c}
\displaystyle (3/4) A + A_{\theta\theta} + A_\theta \cot{\theta} = B, \\[1.5ex]
\displaystyle (25/16) C^2 + C_\theta^2 =  A, \\[1.5ex]
\displaystyle (15/16) BC + B_\theta C_\theta + BC_{\theta\theta} 
   + BC_\theta \cot{\theta} = 0.
\end{array}
\label{sysabc}
\end{equation}
Note that the last equation can be simplified by introducing an auxiliary 
  function $D(\theta) \equiv BC_\theta \sin{\theta}$:
\begin{equation}
D_\theta C_\theta = - (15/16) DC.
\label{sysf}
\end{equation}

Solutions of Eqs. (\ref{sysabc}) must satisfy a number of conditions at 
  the symmetry axis ($\theta=0$) and on the cone surface 
  ($\theta=\theta_m\equiv \pi-\alpha$): 
\begin{equation}
A(0)=1, \quad C_\theta(0) = 0, \quad A(\theta_m) = 0.
\label{gy}
\end{equation}
Note that the condition that the initial velocities of emitted particles are 
  equal to zero, i.e., the pair of conditions $C(\theta_m)=0$ and 
  $C_\theta(\theta_m)=0$, follows from the second equation of the set (\ref{sysabc}).
Differentiating the same equation with respect to $\theta$, we find that 
  the condition $A_\theta(0)=0$ is satisfied if the second derivative of the 
  function $C(\theta)$ is finite at the symmetry axis. 
So, the solution of the problem (\ref{sysabc}) and (\ref{gy}) is unique for a given
  value of $\theta_m$.

The additional boundary condition  
\begin{equation}
A_\theta(\theta_m) = -V^{-1}\sqrt{\cot{\alpha}},
\end{equation}
  resulting from the pressure balance condition (\ref{eqdim}), is not used 
  in seeking a solution of Eqs. (\ref{sysabc}). 
It allows us to determine the value of the parameter $V$ that corresponds
  to the angle $\theta_m$. 
Now, let us rewrite this condition in the following form:
\begin{equation}
V = \sqrt{\cot\alpha}/E(\alpha), \qquad  E(\alpha) \equiv - A_\theta(\theta_m).
\label{eqdimA}
\end{equation}
The function $E(\alpha)$ defines the dimensionless value of the electric field
  strength on the cone surface at a distance $r=1$ from the apex 
  ($E(\alpha) = \left|\nabla\phi\right|_{r=1,\,\theta=\theta_m}$).  

The intensity of the ion flux can be characterized by the electrical current
  $I$ (in dimensional form) flowing through the sphere of radius $L$
  with a center at the apex of the cone:
\begin{equation}
I=2\pi q L^2\int_0^{\theta_m}\! N \Psi_R \sin{\theta}\, d\theta.
\end{equation}
As a corresponding dimensionless quantity, it is convenient to take the group
\begin{equation}
J\equiv I\,\frac{\sqrt{m}}{(32q^2\varepsilon_0L^3\sigma^3)^{1/4}},
\end{equation}
  which does not contain the potential difference $\Phi_L$. 
It can be expressed in terms of the functions $B$ and $C$ by following relation:
\begin{equation}
J = V^{3/2} F(\alpha), \qquad
F(\alpha) \equiv \frac{5\pi}{2} \int_0^{\theta_m}\! BC \sin{\theta}\, d\theta.
\label{Jr}
\end{equation}
The function $F(\alpha)$ defines the particle flux from the top part
  of the cone, $0<r<1$.

The self-similar solutions (\ref{rasdel}) give the ion flux density 
  ($j\sim n|\nabla\psi|$) and the electric field intensity
  ($|\nabla\phi|$) proportional to $r^{-5/4}$ and, respectively,
  to $r^{-1/2}$.
This corresponds to the power law between the quantities $j$ and $|\nabla\phi|$
  on the cone surface, 
\begin{equation}
j\sim |\nabla\phi|^{5/2}.
\label{modellaw}
\end{equation}
It correctly reflects the basic property of the system, namely, the nonlinear
  growth of current density with increasing intensity.
Nevertheless, the $5/2$ power law certainly differs from the exponential
  dependence determined by the kinetics of the field evaporation process
  \cite{l.Muller}. 

Our solutions would be exact if the actual relation between the flux density
  and the electric field were the same as the model relation (\ref{modellaw}). 
Note that the self-similar solutions corresponding to the model law
  of emission (\ref{modellaw}) are the only solutions consistent with our
  main assumption that the surface of a liquid metal is conical.
Any distinction of the law of emission from the model law (\ref{modellaw})
  will lead to deviation of the surface from the ideal cone.

However, it is known from the experiments
  \cite{l.In,l.Ga,l.AuSi,l.Sn,l.CoGe,l.AuGe} that the surface takes
  a near-conical shape, and it is possible to associate it with a certain
  cone half-angle $\alpha$ (the method of angle measurement is given
  in the above-mentioned papers).
That is, the details of the current density distribution over the emitter
  surface (basically, ions evaporate from the protrusion growing at the
  cone apex) have a relatively small influence on the balance between
  the electrostatic and capillary forces at the periphery of the cone structures.
The reason is that the influence of the space charge has an integral character
  (this phenomenon is most conspicuous in planar geometry, where the screening
  effect of the space charge in principle does not depend on its distribution).
As a consequence, our approach, not claiming to describe LMIS operation in detail,
  is rather applicable for the analysis of the integral (averaged)
  characteristics of LMIS, including the relations between the cone half-angle
  $\alpha$, the total emission current $J$, and the applied voltage $V$.
The relations between these main model parameters is determined by the
  expressions (\ref{eqdimA}) and (\ref{Jr}).
In the next section, the set of Eqs. (\ref{sysabc}) will be solved numerically. 
It will allow us to find the auxiliary functions $E(\alpha)$ and $F(\alpha)$,
  which appeared in (\ref{eqdimA}) and (\ref{Jr}). 
As a consequence, the dependence of $\alpha$ on $J$ ($V$ plays the role of 
  the parameter) and the current-voltage dependence ($\alpha$ is the parameter)
  will be established.

\section{Construction of solutions}

\subsection{Asymptotic expansions}

In order to solve the ordinary differential equations (\ref{sysabc}) and 
  (\ref{sysf}) with the conditions (\ref{gy}) numerically, we should use
  asymptotic expansions for the unknown functions at $\theta\to 0$ and
  $\theta\to\theta_m$. 
This is caused by the singular behavior of the functions (or their derivatives)
  at the boundaries $\theta=0$ and $\theta=\theta_m$. 
As will be shown below, $B\to\infty$ at $\theta\to\theta_m$ and 
  $B_{\theta\theta\theta}\to\infty$ at $\theta\to 0$.

At $\theta\to 0$, i.e., at the symmetry axis, the functions $A(\theta)$, 
  $B(\theta)$, $C(\theta)$, and $D(\theta)$ can be expanded into the series
\begin{equation}
\begin{array}{l}
\displaystyle A = 1 + a_2 \theta^{2} + a_4 \theta^{4} + \dots, \\[1.5ex]
\displaystyle B = b_0 \theta^{\beta} + b_0 b_1 \theta^{2+\beta} + \dots, \\[1.5ex]
\displaystyle C = 4/5 + c_2 \theta^{2} + c_4 \theta^{2} + \dots, \\[1.5ex]
\displaystyle D = d_0 \theta^{2+\beta} + d_0 d_2 \theta^{4+\beta} + \dots.
\end{array}
\label{ass0}
\end{equation}
Substitution of these expressions into the initial equations (\ref{sysabc}),
  (\ref{sysf}) yields
\begin{equation}
\begin{array}{lll}
\displaystyle a_2 = - \frac{3}{2^4}, & \displaystyle a_4 = \frac{1}{2^{10}},
  & \dots, \\[1.5ex]
\displaystyle c_2 = \frac{-5+\sqrt{13}}{2^4}, &
  \displaystyle c_4 = \frac{295-107\sqrt{13}}{9\cdot 2^{10}}, & \dots, \\[1.5ex]
\displaystyle \beta = \frac{\sqrt{13}+1}{2} \simeq 2.3, &
  \displaystyle b_1 = - \frac{275+173\sqrt{13}}{1152}, & \dots, \\[1.5ex]
\displaystyle d_0 = 2 b_0 c_2, & \displaystyle d_2 = - \frac{165+31\sqrt{13}}{384},
  & \displaystyle \dots.
\end{array}
\end{equation}
The expansions (\ref{ass0}) satisfy two first conditions from (\ref{gy}); 
  they contains a free parameter (the coefficient $b_0$), which is determined by 
  the following condition at the cone surface: $A(\theta_m)=0$. 
Note that the divergence of higher derivatives of the function $B$ leads to the
  divergence of higher derivatives of the functions $A$ and $C$. 
The next terms of the expansions (\ref{ass0}) for $A$ and $C$ are of the 
  order of $\theta^{2+\beta}$.

In the limit $\theta\to\theta_m$, i.e., on the cone surface, the unknown functions 
  can be expanded into power series in the parameter $x=\theta_m-\theta$:
\begin{equation}
\begin{array}{l}
A = x^{1/2} \sum\limits_{i=1} a'_i x^{i/2}, \ \ \ \ \
B = x^{-1} \sum\limits_{i=1} b'_i x^{i/2}, \\ [2.0ex]
C = x \sum\limits_{i=1} c'_i x^{i/2}, \ \ 
D = d'_0 \left( 1 + x^{3/2} \sum\limits_{i=1} d'_i x^{i/2} \right).
\end{array}
\label{asstm}
\end{equation}
The first coefficients of these expansions are listed below:
\begin{equation}
\begin{array}{c}
\displaystyle  a'_2 = \frac43\, b'_1, \ \ \ \  c'_1 = \frac23 \,\sqrt{a'_1}, \\[2.0ex]
\displaystyle  c'_2 = \frac13 \,\frac{b'_1}{\sqrt{a'_1}}, \ \ \ \
  b'_2 = -\frac23 \,\frac{b'_1{}^2}{a'_1}, \\[2.0ex]
\displaystyle d'_0 = -b'_1 \sqrt{a'_1} \sin{\theta_m}, \ \ \ \ d'_1 = - \frac{5}{16}.
\end{array}
\end{equation}
Free parameters of these expansions, namely, the coefficients $a'_1$ and $b'_1$,
  are determined by the boundary conditions at the symmetry axis. 

Increasing the charge density over the cone leads to screening of the electric
  field at its surface, i.e., the coefficient $a'_1$ will approach zero.
In the formal limit $a'_1 = 0$ (the electric field turns to zero at the anode surface),
  another asymptotics is realized:
\begin{equation}
\begin{array}{c}
A = x^{1/3} \sum\limits_{i=1} a''_i x^i, \ \ \ \ 
B = x^{-5/3} \sum\limits_{i=1} b''_i x^i, \\[2.0ex]
C = x^{2/3} \sum\limits_{i=1} c''_i x^i, \ \ \ \
D = d''_0 \left( 1 + x \sum\limits_{i=1} d''_i x^i \right),
\end{array}
\label{asstm0}
\end{equation}
  where
\begin{equation}
\begin{array}{c}
\displaystyle a''_1 = \frac94 \, b''_1, \ \ \ \  
  a''_2 = \frac{6}{5} \, b''_1 \cot{\theta_c}, \\[2.0ex]
\displaystyle c''_1 = \frac{9}{10} \,\sqrt{b''_1}, \ \ \ \
  c''_2 = \frac{3}{20} \,\sqrt{b''_1} \cot{\theta_c}, \\[2.0ex]
\displaystyle b''_2 = \frac{11}{15} \,b''_1 \cot{\theta_c}, \ \ \ \
  d''_0 = -\frac32\, b''_1{}^{3/2} \sin{\theta_c}, \\[2.0ex]
\displaystyle  d''_1 = - \frac{9}{32}, \ \ \ \cdots.
\end{array}
\end{equation}
These expressions contain two free parameters: the coefficient $b''_1$
  (or $d''_0$) and the angle $\theta_m\equiv\theta_c$ at which complete
  screening of the external field occurs.

It should be noted that the function $D$ converges rapidly near the cone
  surface for both forms of expansion, (\ref{asstm}) and (\ref{asstm0}). 
Therefore, using the function $D$ is preferable to using the function $B$
  in the procedure of numerical integration of the problem
  (\ref{sysabc})--(\ref{gy}).

\begin{figure}
\centering
\includegraphics*{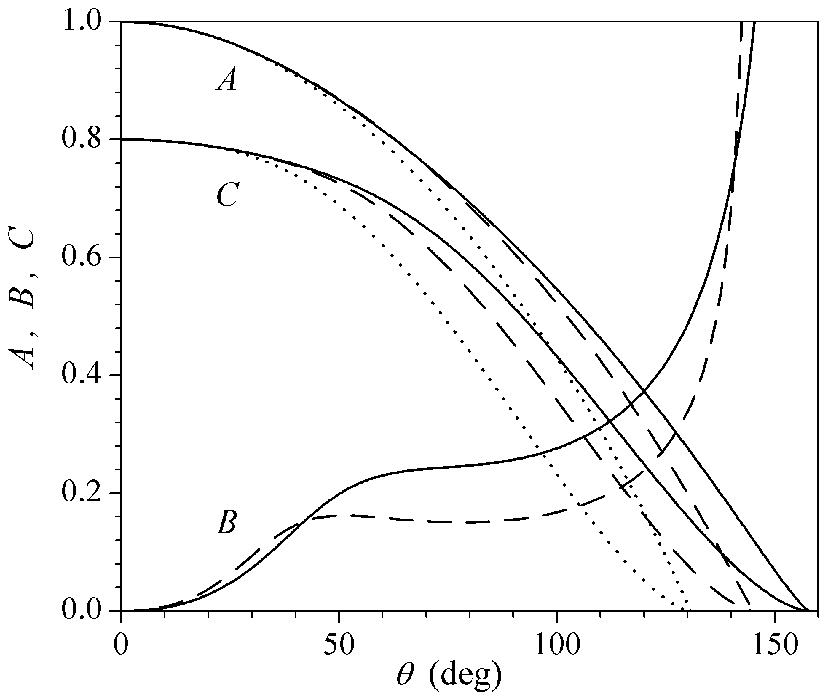}
\caption{Solutions of the system (\ref{sysabc}) for the cone half-angles
  $\alpha = \alpha_T$ (in this case, $B\equiv 0$) (dotted lines), $\alpha=35^\circ$
  (dashed lines), and $\alpha = \alpha_c$ (solid lines).}
\label{f.Gabc}
\end{figure}

\subsection{Numerical calculations}

The set of Eqs. (\ref{sysabc}), (\ref{sysf}) was solved numerically 
  by the prediction-correction method of the third order. 
The calculation starts from the asymptotics (\ref{asstm}) or (\ref{asstm0}).
It has been found that the equations admit solutions for the 
  angles in the range $\theta_T\leq\theta_m\leq\theta_c$. 
The minimum angle value $\theta_T\simeq 130.71^{\circ}$ corresponds to 
  the Taylor cone ($\alpha=\alpha_T\simeq 49.29^{\circ}$) and refers to 
  the special case of no space charge, i.e., $B=0$. 
For this case, the $\theta$ dependence of $A$ is determined by the Legendre
  function $P_{1/2}(\cos{\theta})$. 
The upper bound of the angle, $\theta_c$, equals $\simeq 158.11^{\circ}$. 
It corresponds to the least possible half-angle of the cone 
  $\alpha=\alpha_c\simeq 21.89^{\circ}$. 
This configuration of the surface relates to the formal limit that the 
  electric field at the cone is completely screened by the space charge (this
  limit cannot be achieved because of the finite emissivity of the surface). 
The results of the calculations, corresponding to different values of $\alpha$,
  are presented in Fig.~\ref{f.Gabc}. 
In view of the obvious relations
\begin{equation}
\begin{array}{l}
\nabla\phi = r^{-1/2} \left\{ A/2,A_\theta,0 \right\}, \\[2.0ex]
\nabla\psi = r^{1/4} \left\{ 5C/4,C_\theta,0 \right\},
\end{array}
\label{Ed}
\end{equation}
  the angle dependence of $A(\theta)$ and $C(\theta)$ give an idea of the 
  distributions of the electric field and the velocity field. 
As an example, Fig.~\ref{f.G01} shows the equipotential surfaces and streamlines
  of the flow corresponding to $\alpha = 45^{\circ}$.

\begin{figure}
\centering
\includegraphics*{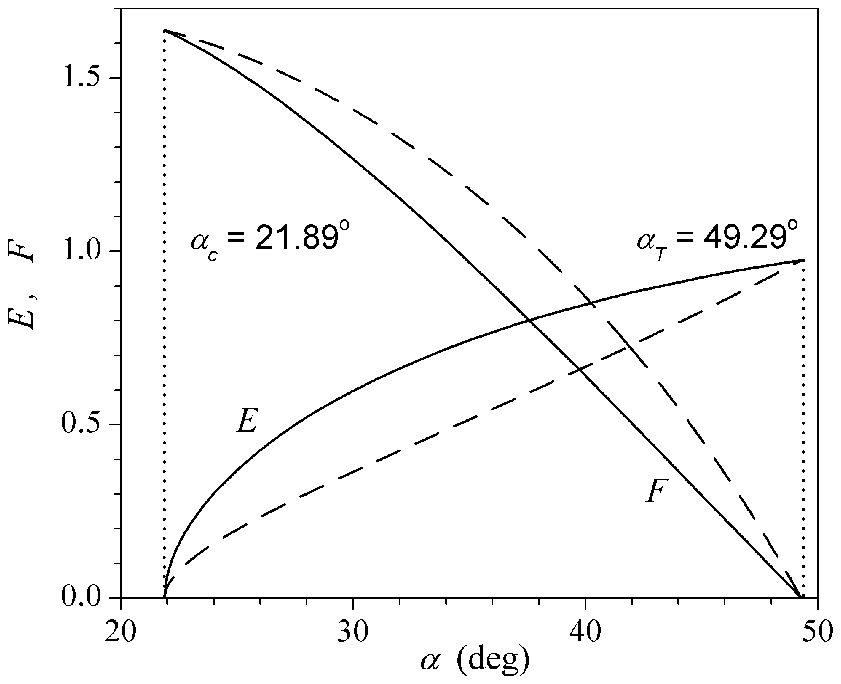}
\caption{The auxiliary functions $E(\alpha)$ and $F(\alpha)$ (solid lines), 
  which represent the electric field strength on the cone surface at $r=1$
  and, respectively, the ion flux from the top part of the cone ($0<r<1$).
The dashed lines correspond to the one-dimensional model
  (Eqs.~(\ref{onemod})--(\ref{plane})).}
\label{f.Gec}
\end{figure}

\subsection{The electric field strength at the cone surface}

The calculations show that the function $E(\alpha)$ grows monotonically 
  from zero to the value $\simeq 0.975$ as the angle $\alpha$ increases
  from $\alpha_c$ to $\alpha_T$ (see Fig.~\ref{f.Gec}). 
Near the limiting case (\ref{asstm0}), the estimate 
  $E(\alpha)\sim(\alpha-\alpha_c)^{1/2}$ is valid for $\alpha\to\alpha_c$. 
To verify this dependence, it is possible to draw an analogy between the
  considered problem and the problem of a charged-particle flow in a plane
  vacuum diode, which can be solved analytically. 
Actually, in the one-dimensional case, where all quantities depend only on 
  one coordinate $z$, the equations (\ref{sysd}) reduce to the set of ordinary
  differential equations
\begin{equation}
\phi_{zz}=n, \quad {\psi_z}^2=\phi, \quad (n\psi_z)_z=0, 
\label{planar}
\end{equation}
  which look like the equations (\ref{sysabc}) after the substitution $z\to\theta$. 
These equations admit the following exact implicit solution for the electric
  field potential $\phi$:
\begin{equation}
6j^2z - {E_0}^3 = \left(2j\sqrt{\phi} - {E_0}^2\right) \sqrt{4j\sqrt{\phi} + {E_0}^2},
\label{onecase}
\end{equation}
  where $j=n\psi_z$ is the constant current density, $E_0=\phi_z|_{z=0}$ is
  the electric field strength at the planar emitter. 
Let us also introduce the notation for the electric strength at the opposite
  electrode, $E_1=\phi_z|_{z=h}$, ($h$ is the interelectrode distance) and
  for the electric potential on it, $\phi_h=\phi|_{z=h}$. 
From the solution (\ref{onecase}) one can obtain the relations between the 
  quantities $E_0$, $E_1$, $\phi_h$, $h$, and $j$,  
\begin{equation}
\begin{array}{c}
\displaystyle 4j\sqrt{\phi_h}={E_1}^2 - {E_0}^2, \\[1.5ex]
\displaystyle 6j^2h - {E_0}^3 = \left(2j\sqrt{\phi_h} - {E_0}^2\right) E_1~. 
\end{array}
\label{onem}
\end{equation}
By analogy with $\theta_c$, we introduce the interelectrode distance
  $h_c=4\phi_h/(3E_1)$, corresponding to the limit $E_0=0$,
  and then rewrite the expressions (\ref{onem}) as
\begin{equation}
j = \frac{4\phi_h^{3/2}}{9h_c^2} \left( 1 - \frac{E_0^2}{E_1^2} \right), \ \ \ \ 
\frac{h}{h_c} = 1 - \left( \frac{E_0}{E_1+E_0} \right)^2.
\label{onemod}
\end{equation}
Note that, in the limit $E_0/E_1\to 0$, the first equation of (\ref{onemod})
  represents the Child-Langmuir law \cite{l.Child,l.Langmuir} for a plane vacuum diode.
It follows from the second equation of (\ref{onemod}) that
  $E_0 \approx E_1\sqrt{1-h/h_c}$ in the same limit,
  i.e., for fixed $\phi_h$ and $E_1$, the electric field strength at the emitter
  surface $E_0$ has a square root dependence on the small quantity $h_c-h$. 
For $h=h_c$, the electric field at the emitter surface is completely screened 
  by the space charge. 
At given $\phi_h$ and $E_1$, the space charge density decreases as the interelectrode
  distance $h$ is reduced. 
The opposite limit, where the space charge is absent ($E_0 = E_1$ and $\phi(z) = E_1 z$),
  is reached at $h=(3/4)h_c$.

\begin{figure}
\centering
\includegraphics*{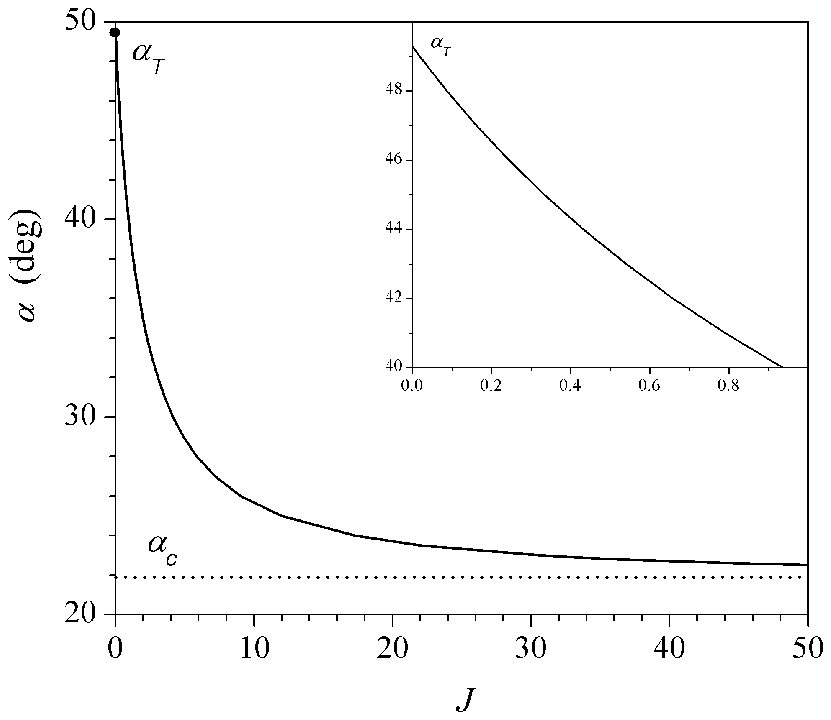}
\caption{Dependence of the cone half-angle $\alpha$ on the dimensionless
  emission current $J$. The inset demonstrates the same dependence for
  relatively small currents. } 
\label{f.fig_03}
\end{figure}

So, in the framework of the discussed analogy we can identify 
  (i) the angles $\theta_m$, $\theta_c$, and $\theta_T$ with 
  the interelectrode distances $h$, $h_c$, and $h_T = (3/4)h_c$, respectively, 
  (ii) the electric field strength on the cone surface at a unit distance from
  the apex $E$ with the strength at the surface of the planar emitter surface $E_0$, 
  (iii) the electric field strength at the symmetry axis at a unit distance from
  the cone apex $|\nabla\phi|=1/2$ with the strength at the opposite electrode $E_1$,
  and also (iv) unit potential difference between the cone axis and its surface
  for $r=1$ with the fixed potential difference $\phi_h$ for a plane vacuum diode. 
In order to compare the characteristics of conical and planar diodes, we apply
  the linear mapping of the angle interval $\theta_T\leq\theta_m\leq\theta_c$
  into the interval of distances $h_T\leq h\leq h_c$,
\begin{equation}
\frac{h-h_T}{h-h_c} = \frac{\theta_m-\theta_c}{\theta_m-\theta_T}.
\label{htheta}
\end{equation}
The analogs of the functions $F(\alpha)$ and $E(\alpha)$ corresponding to the
  one-dimensional model (\ref{planar}), 
\begin{equation}
F_{pl} = \frac{9j\,h_c^2} {4\phi_h^{3/2}} \, F(\alpha_c)~, \ \ \ \ \
E_{pl} = \frac{E_0}{E_1} \ E(\alpha_T)~,
\label{plane}
\end{equation}
  are presented in Fig.~\ref{f.Gec}. 
It can be seen that the properties of conical (\ref{sysabc}) and planar 
  (\ref{planar}) models are qualitatively similar. 
Consequently, we should expect that $E\sim(\theta_c-\theta_m)^{1/2}$ at
  $\theta_m\to\theta_c$, or, in terms of the cone half-angle, 
  $E\sim (\alpha-\alpha_c)^{1/2}$ at $\alpha\to\alpha_c$, as we wished to show.

Taking into account such a dependence near the limiting regime (\ref{asstm0}),
  we approximate the calculated dependence of $E$ on $\alpha$, by the formula 
\begin{equation}
E(\alpha) = - 0.9754 \, \xi - 0.6176 \, \xi^2 + 0.8342 \, \xi^3 -0.2164 \, \xi^4,
\label{apprec}
\end{equation}
  where $\xi = \left[ (\alpha-\alpha_c)/(\alpha_T-\alpha_c) \right]^{1/2}$. 
The error of the approximation (\ref{apprec}) is less than 0.1\% over the
  range $\alpha_c<\alpha<\alpha_T$. 
According to the balance condition (\ref{eqdimA}), the obtained dependence
  $E(\alpha)$ determines the relation between geometry of the emitting cusp,
  i.e., the angle $\alpha$, and the external control parameter of the system,
  i.e., the potential $V\sim\Phi_L$.

\begin{figure}
\centering
\includegraphics*{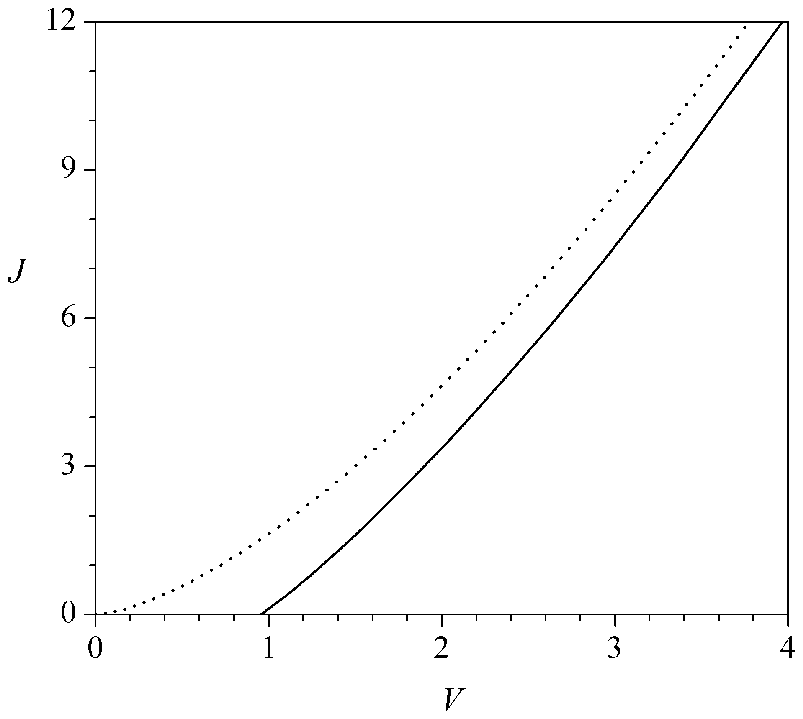}
\caption{The current $J$ as a function of the applied potential difference $V$.
  The dotted line shows the asymptotic shape of the current-voltage
  characteristic corresponding to the Child-Langmuir law. }
\label{f.GVA}
\end{figure}

\subsection{The relations between model parameters}

\begin{figure*}
\centering
\includegraphics*{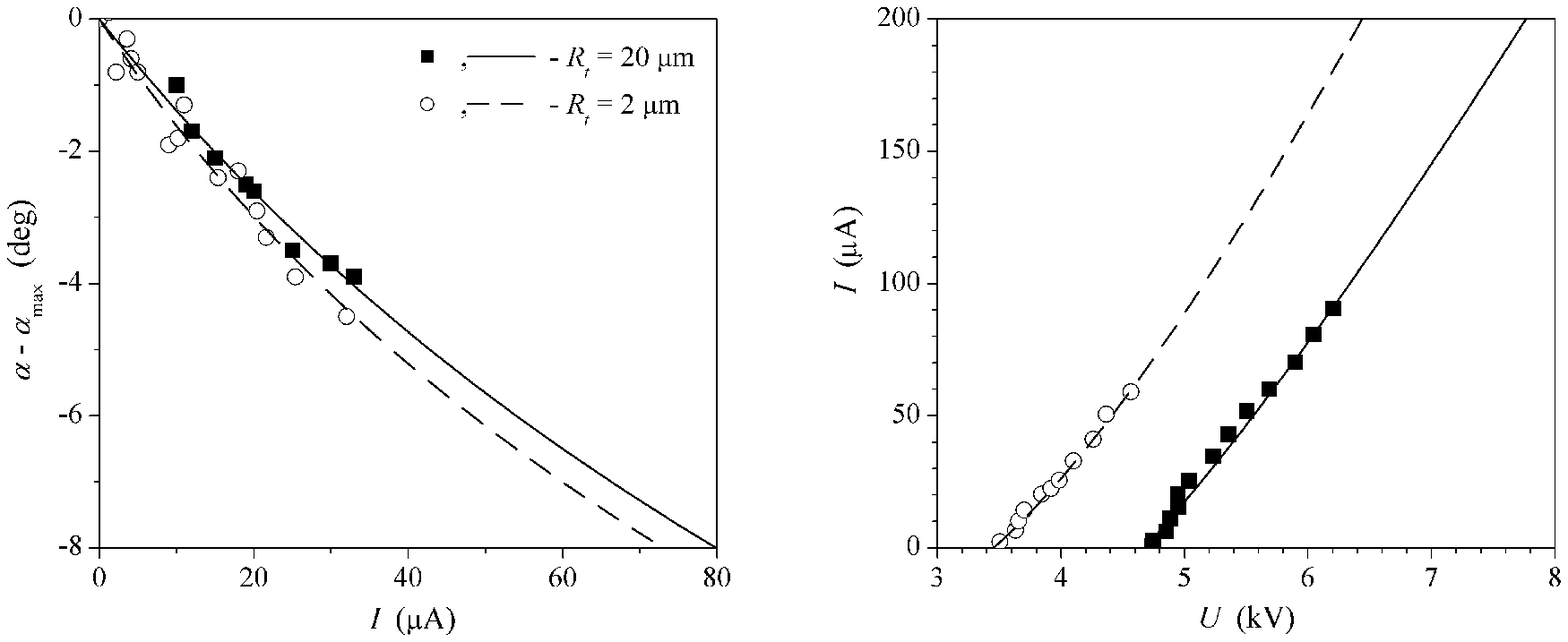}
\caption{The change in the cone half-angle versus the emission current (left)
  and the current-voltage characteristics (right) for a gallium liquid-metal
  ion source. The points correspond to the experimental data of \cite{l.Ga}
  and the lines to the theory.}
\label{f.GGa}
\end{figure*}

The quantity $F(\alpha)$, which is determined by (\ref{Jr}), specifies a 
  particle flux from the top part of the cone, $0<r<1$. 
The angular dependence of this function obtained as a result of numerical
  solution of the problem (\ref{sysabc})--(\ref{gy}) is presented
  in Fig.~\ref{f.Gec}. 
The calculated function $F(\alpha)$ can be approximated by the relation
\begin{equation}
F(\alpha) \simeq 1.8863\zeta+ 0.0956\zeta ^2 - 0.2259\zeta ^3 - 0.1167\zeta ^4,
\label{apprJ}
\end{equation}
  where $\zeta(\alpha) = 1-\xi^2 =(\alpha_T-\alpha)/(\alpha_T-\alpha_c)$.
The error of this approximation is less than 0.15\%. 
Note that the function $F(\alpha)$ monotonically decreases from $\simeq 1.637$
  to zero as the angle $\alpha$ changes from $\alpha_c$ to $\alpha_T$. 

The expressions (\ref{eqdimA}) and (\ref{Jr}) together with (\ref{apprec})
  and (\ref{apprJ}) allow us to obtain the dependence of the cone half-angle
  $\alpha$ on the current $J$ as well as the current-voltage characteristic
  of the cone, i.e., the dependence of $J$ on $V$. 
These relationships are plotted in Figs.~\ref{f.fig_03} and \ref{f.GVA}, respectively.

From Fig.~\ref{f.fig_03} it can be seen that the angle monotonically decreases
  with increasing current. 
It is equal to the Taylor angle $\alpha_T\approx 49.29^{\circ}$ at zero current
  and tends to the angle $\alpha_c\approx 21.89^{\circ}$ in the formal limit of
  an infinite current. 
Note that the interpretation of experiments 
  \cite{l.In,l.Ga,l.AuSi,l.Sn,l.CoGe,l.AuGe}
  in the framework of our model corresponds to the dimensionless current
  in the range $0<J<1$ (see the inset of Fig. \ref{f.fig_03}). 
For larger current values, the cone structure becomes instable.

An important feature of the theoretical current-voltage characteristic
  (Fig.~\ref{f.GVA}) is its threshold character. 
There is no electric current, $J=0$, if 
  $V<V_0=\sqrt{\mbox{ctg}\,\alpha_T}/f(\alpha_T)\simeq 0.9512$. 
This is related to the impossibility of a balance between the electrostatic
  and capillary forces at a relatively small potential difference. 
The capillary force will dominate and the cone structure will break. 
In the formal limit of large $V$ the model yields the universal Child-Langmuir
  law: $J\to F(\alpha_c)V^{3/2}$, which describes the regime of space-charge
  limitation of the current as a result of complete screening of the electric
  field at the electrode surface.

Excluding the parameter $\alpha$ from (\ref{eqdimA}) and (\ref{Jr}), we can
  rewrite the dependence of $J$ on $V$ in the explicit form. 
The function $F$ linearly goes up to the value $F(\alpha_c)$ as 
  $\alpha\to\alpha_c$ (see Fig.~\ref{f.Gec}), and, as discussed above, 
  the function $E$ tends to zero according to the square root law. 
As a consequence, the dependence of $J$ on $V$ must have the form
  $J\approx j_0V^{3/2}+j_1 V^{-1/2}$ in the limit $V\to\infty$. 
With this estimate taken into account, for $V>V_0$ the required dependence 
  can be approximated by the expression
\begin{equation}
J = V^{3/2}\left(j_0+\frac{j_1}{V^2}+\frac{j_2}{V^4}+\frac{j_3}{V^6}\right).
\label{apprJV}
\end{equation}
The coefficients of the approximation (\ref{apprJV}) are the following: 
  $j_0 = 1.6372$, $j_1 = -1.8635$, $j_2 = 0.3523$, $j_3 = -0.00676$; 
  the approximation error is less than 0.00015 in the absolute value.

Thus, the relations (\ref{eqdimA}) and (\ref{Jr}) together with the approximate
  expressions (\ref{apprec}), (\ref{apprJ}), and (\ref{apprJV})
  completely determine the integral characteristics of the system. 
They establish relationships between the following quantities: the applied potential
  difference $V$, the emission current $J$, and the cone half-angle $\alpha$. 
This will allow us to compare our theoretical calculations with available
  experimental data in the next Section.

\section{Comparison with experimental data}

\begin{figure*}
\centering
\includegraphics*{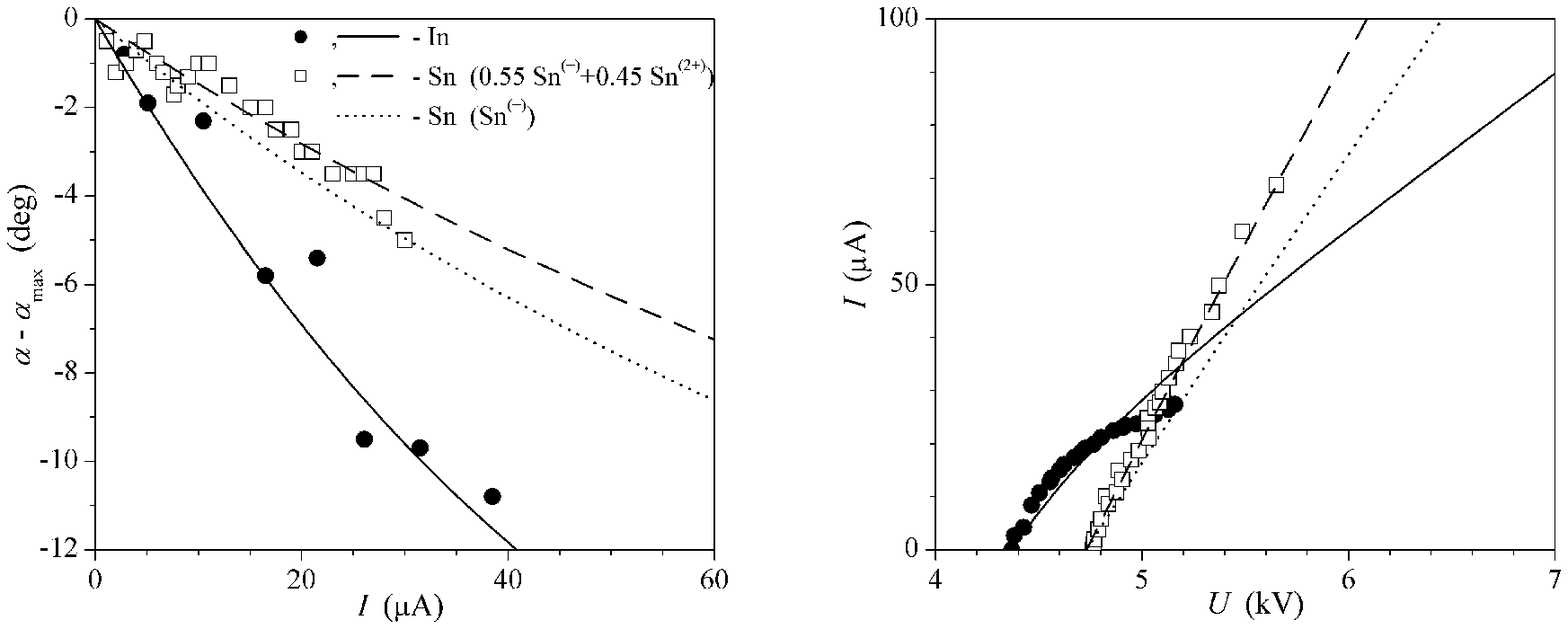}
\caption{The change in the cone half-angle versus the emission current
  (left) and the current-voltage characteristics (right) for indium and
  tin liquid-metal sources. The points correspond to the experimental
  data of \cite{l.In,l.Sn} and the lines to the theory;
  the solid line corresponds to indium, the dashed line to tin, and 
  the dotted line to tin under the condition of the emission of
  singly-charged ions Sn${}^{(+)}$ only. }
\label{f.InSn}
\end{figure*}

In order to compare the results of our calculations with experimental data, 
  we assume that the control parameter of the model $\Phi_L$ is directly
  proportional to the potential difference $U$ applied to the interelectrode
  space, i.e.,
\begin{equation}
\Phi_L = \kappa U, 
\end{equation}
  where the proportionality coefficient $\kappa$ is determined by 
  constructional features of an experimental facility and does not depend
  on the emission current. 
Then, after switching to dimensional quantities, the expressions
  (\ref{eqdimA}) and (\ref{Jr}) become
\begin{equation}
U = \left(\frac{2\sigma}{\varepsilon_0}\right)^{1/2} 
  \frac{\sqrt{L(\alpha)\cot{\alpha}}}{\kappa E(\alpha)},
\label{Valf}
\end{equation}
\begin{equation}
I = \varepsilon_0 \left(\frac{2q}{m}\right)^{1/2}\! 
  F(\alpha)\left(\kappa U\right)^{3/2}.
\label{IV}
\end{equation}
Here we take into account that the characteristic size $L$ of the top part
  of the infinite model cone (the electric current from this part of the 
  cone is identified with the emission current from the experimentally
  observed conical spike) can depend on the value of the emission current
  and, as a consequence, on the angle $\alpha$. 

The relations (\ref{Valf}) and (\ref{IV}) together with the approximate
  formulas (\ref{apprec}), (\ref{apprJ}) or (\ref{apprec}), (\ref{apprJV})
  allow one to determine the free parameters of the model,
  namely, the characteristic size of the cone $L(\alpha)$ and the coefficient
  $\kappa$ characterizing the electric field distribution in the experimental
  facilities, from the available experimental data on current-voltage and
  current-angle dependencies for liquid-metal ion sources. 
We have used data for emission into vacuum from liquid indium \cite{l.In},
  gallium \cite{l.Ga}, tin \cite{l.Sn}, and also from liquid alloys Au$+$Si
  \cite{l.AuSi}, Co$+$Ge \cite{l.CoGe}, and Au$+$Ge \cite{l.AuGe}. 
In our calculations we have taken the following values of the surface tension:
  $\sigma = 0.572$ N/m (In, \cite{l.Sn}), 0.735 N/m (Ga, \cite{l.Sn}),
  0.560 N/m (Sn, \cite{l.Sn}), 2.20 N/m (Au$+$Si, \cite{l.AuGe}),
  1.84 N/m (Co+Ge, \cite{l.CoGe}), and 1.62 N/m (Au$+$Ge, \cite{l.AuGe}).
The angular dependence of $L$ was approximated by the two-parameter function 
\begin{equation}
L(\alpha)=L_c+\left(L_T-L_c\right)\left(\frac{\alpha-\alpha_c}
  {\alpha_T-\alpha_c}\right)^2,
\label{Lalf}
\end{equation}
  where $L_T$ and $L_c$ are parameters. 
It should be noted that the maximum half-angle $\alpha_{\max}$,
  corresponding to zero emission current in the experiments
  \cite{l.In,l.Ga,l.Sn,l.AuSi,l.CoGe,l.AuGe}, slightly differs from the Taylor
  angle $\alpha_T$. 
Then, for comparison with the theory, the experimentally observed values 
  of $\alpha$ have been corrected by $\alpha_T-\alpha_{\max}$.

\begin{figure*}
\centering
\includegraphics*{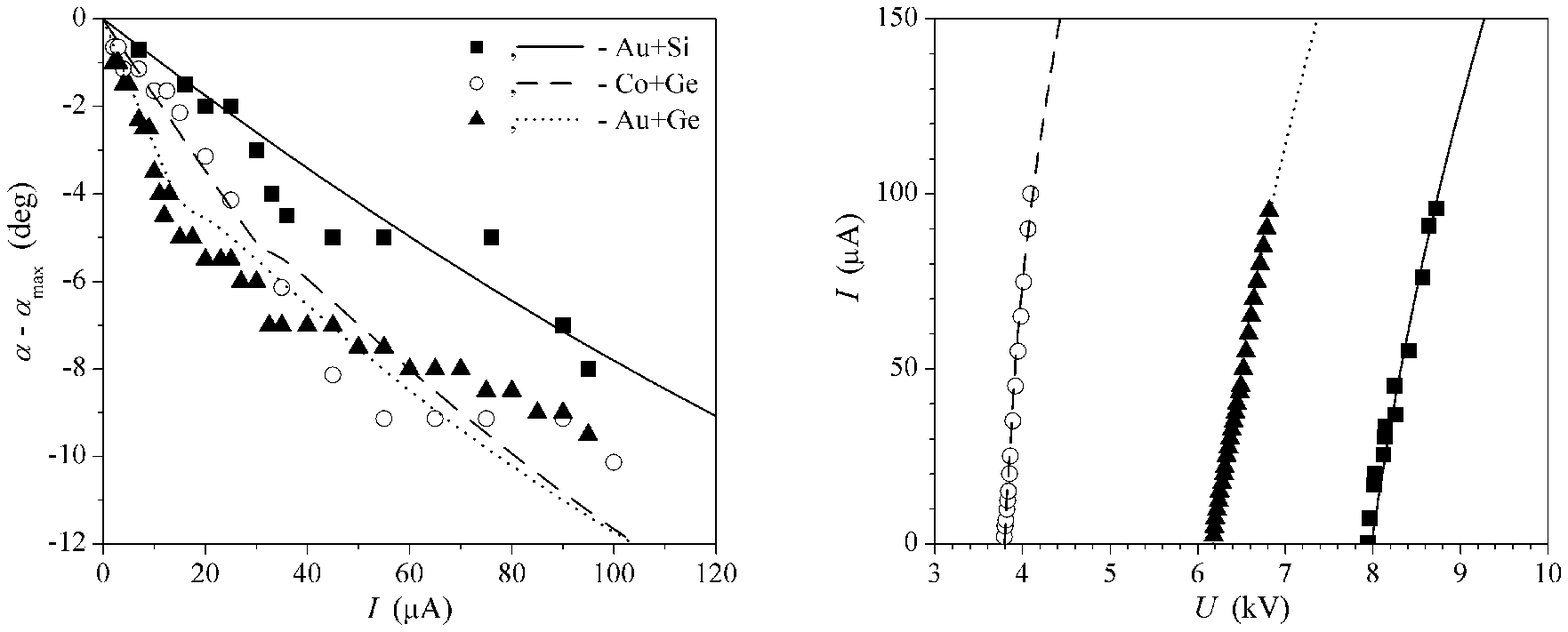}
\caption{The change in the cone half-angle versus the emission current
  (left) and the current-voltage characteristics (right) for Au$+$Si
  \cite{l.AuSi}, Co$+$Ge \cite{l.CoGe}, and Au$+$Ge \cite{l.AuGe} alloy
  liquid-metal ion sources. The points correspond to the experimental
  data and the lines to the theory.}
\label{f.mix}
\end{figure*}

Due to its physical properties (low temperature of melting, low pressure
  of saturated vapor, primarily single ionization in the emission processes),
  gallium is the most convenient and reliable metal for investigating
  the operation of liquid-metal ion sources \cite{l.Forbes}. 
In the work \cite{l.Ga} the gallium ion sources, including a tungsten needles
  with tip radius $R_t=2$~$\mu$m and $R_t=20$~$\mu$m, were investigated. 
As a result of treating data on the current-voltage characteristics and 
  geometry of the observed structures on the surface of liquid gallium,
  presented in \cite{l.Ga}, we obtain: $\kappa=0.0905$, $L_T=0.65$~$\mu$m,
  $L_c=0.84$~$\mu$m ($R_t=2$~$\mu$m); $\kappa=0.076$,  $L_T=0.84$~$\mu$m,
  $L_c=0.84$~$\mu$m ($R_t=20$~$\mu$m).
The achieved quality of describing the experimental data is demonstrated
  in Fig.~\ref{f.GGa}. 
Application of a sharper needle (with $R_t=2$~$\mu$m) provides a stronger
  focusing of the electric field that corresponds to a larger value of 
  the coefficient $\kappa$ and, consequently, to a smaller value of the 
  threshold potential difference (see Fig.~\ref{f.GGa}). 
On the other hand, a small initial radius of curvature probably begins
  to ``constrict'' the developing cone; the value of $L_T$ corresponding
  to $R_t=2$~$\mu$m is somewhat smaller than for $R_t=20$~$\mu$m. 
According to the relations (\ref{Valf}) and (\ref{IV}), smaller size of the
  cone leads to faster change of the cone half-angle with increase in 
  emission current, all other parameters (the surface tension coefficient,
  the mass and charge of emitted particles) being the same (see Fig. \ref{f.GGa}). 
Note that the obtained values of $L_T$ are comparable with the size of the
  experimentally observed cones (they are of the order of 1~$\mu$m).

Fig.~\ref{f.InSn} shows the current-voltage and current-angle dependencies
  for the conical spikes observed at the surface of liquid indium \cite{l.In}
  and tin \cite{l.Sn}. 
Probably, the strongly nonlinear character of the current-voltage dependence
  for indium is an experimental error \cite{l.In}. 
In later publications \cite{l.CoGe,l.AuGe} the same authors approximate
  this dependence by a linear function; its slope is approximately the same
  as for the corresponding theoretical curve in Fig.~\ref{f.InSn}.

It has been noted in \cite{l.Sn} that both singly-charged ions Sn${}^{(+)}$
  and doubly-charged ions Sn${}^{(2+)}$ are emitted from the surface of tin. 
The fraction of doubly-charged ions is $x_2 = N_2/(N_1+N_2) \simeq 45$\%
  \cite{l.AuGe}. 
It is easy to verify that the relations obtained above are applicable for
  a flow consisting of $k$ different types of particles. 
The only necessary modification refers to the expression for total 
  current (\ref{IV}). 
We should apply the following change:
\begin{equation}
\sqrt{\frac{2q}{m}} \   \to \   \sum_{i=1}^{k} \frac{q_i x_i}{Q} 
   \sqrt{\frac{2q_i}{m_i}}, \qquad   Q = \sum_{i=1}^{k} q_i x_i,
\end{equation}
  where $q_i$, $m_i$, and $x_i$ are the charge, mass, and relative fraction
  of particles of $i$-type. 
So, for the case of tin ($q_2=2q_1=2e$, $m_2=m_1=m$), the expression
  (\ref{IV}) transforms into
\begin{equation}
I = \frac{1+(2\sqrt{2}-1)x_2}{1+x_2} \varepsilon_0 
   \left(\frac{2e}{m}\right)^{1/2}\! F(\alpha)\left(\kappa U\right)^{3/2},
\end{equation}
  where $e$ is the elementary charge. 
Indium ions are emitted in the singly-charged state In${}^{(+)}$ \cite{l.AuGe},
  so that such modification is not required for an indium liquid-metal ion source. 
The theoretical curves presented in Fig.~\ref{f.InSn} correspond to the
  following values of the parameters: 
  $\kappa=0.0495$, $L_T=0.4$~$\mu$m, $L_c=0.3$~$\mu$m (In);
  $\kappa=0.075$,  $L_T=1.1$~$\mu$m, $L_c=0.78$~$\mu$m (Sn).

Indium and tin have close atomic weights and surface tension coefficients. 
In this connection, a sharp distinction between the dependences $I(U)$ and
  $\alpha(I)$ corresponding to these metals is surprising \cite{l.Sn}. 
One of the likely reasons of the different behavior of indium and tin is the 
  presence of a large amount of ions Sn$^{2+}$ in the experiments \cite{l.Sn}. 
Ions with larger charge numbers move faster away from the electrode and, 
  as a consequence, provide less screening effect. 
Therefore, an increase in the voltage is accompanied by a larger increase
  in the emission current and by a relatively slow change in the cone shape. 
The dependences $I(U)$ and $\alpha(I)$, corresponding to the assumptions
  that only the singly-charged ions Sn${}^{(+)}$ are emitted from liquid tin
  and the parameters $\kappa$, $L_T$, and $L_c$ are fixed, are shown 
  in Fig.~\ref{f.InSn} by the dotted line. 
One can see that considering only singly-charged ions slightly approaches
  the model characteristics of liquid-metal emitters based on liquid tin and indium.
However, the initial difference between the curves is too large and cannot
  be explained only by the presence of ions Sn${}^{(2+)}$.

Another reason of the behavior difference between liquid-metal ion sources
  is in the conditions of experimental studies \cite{l.In} and \cite{l.Sn}. 
The theoretical model presented in this work points to the essential difference
  between the characteristic sizes $L_T$ of the cones. 
One can assume that the developing cone was ``constricted'' by the small size
  of the tip of a tungsten needle in the experiment with indium 
  ($R_t=1.0$~$\mu$m according to \cite{l.In}). 
Although the value of the tip radius $R_t=4$~$\mu$m for the experiments
  with indium was indicated in the subsequent paper \cite{l.AuGe},
  it was measured with the help of the image of the liquid surface instead
  of the needle. 
The above analysis of gallium emitters with different needles shows that
  the ``constriction'' effect becomes appreciable if $R_t\le 2$~$\mu$m.

Finally, Fig.~\ref{f.mix} shows the dependencies $I(U)$ and $\alpha(I)$
  for liquid alloys Au$+$Si \cite{l.AuSi}, Co$+$Ge \cite{l.CoGe}, 
  and Au$+$Ge \cite{l.AuGe} that correspond to the model coefficients
  $\kappa=0.084$, $L_T=1.0$~$\mu$m, $L_c=0.45$~$\mu$m (Au$+$Si);
  $\kappa=0.084$, $L_T=0.27$~$\mu$m, $L_c=0.08$~$\mu$m (Co$+$Ge);
  $\kappa=0.051$, $L_T=0.30$~$\mu$m, $L_c=0.10$~$\mu$m (Au$+$Ge).
Theory and observations are in rather good agreement only for Au$+$Si
  from the above-listed alloys. 
The alloys containing germanium are distinguished by the presence of
  breaks on the experimental curves $\alpha(I)$. 
Such a behavior cannot be described in the framework of the proposed model. 
The anomalous character of the dependence $\alpha(I)$ can be caused by 
  changing properties of charged particles flow. 
So, ``freezing'' of the cone angle along with the continuing growth of
  the emission current, which is characteristic for the mentioned alloys
  (see Fig.~\ref{f.mix}), allows us to suppose that, from some point, 
  the average ratio of mass and charge dencities over the flow essentially
  decreases. 
Unfortunately, the experimental works \cite{l.CoGe,l.AuGe} does not include
  information about changes in the mass/charge ratio with the emission current. 
In Fig.~\ref{f.mix} the theoretical curves are obtained under the assumption
  that the intensive emission of doubly-charged germanium ions Ge${}^{(2+)}$
  starts with some value of the applied potential difference, $U_1$ 
  (or current, $I_1$). 
The quantities $U_1$, $I_1$ are of the order of $3.87$~kV, 30~mkA for the
  alloy Co$+$Ge, and 6.3~kV, 15~mkA for Au$+$Ge. 
In both cases, the potential drop is $\Phi_L=\kappa U\simeq 0.32$~kV at 
  the distance of the order of the cone size ($R=L$).

\section*{Conclusion}

In the present paper we have developed the self-consistent model describing
  how the space charge near the emitting cone apex affects its shape. 
Our approach is based on the self-similar reduction of the equations,
  which govern the spatial distributions of the electric field, ion velocity
  field, and particle concentration, to the system of ordinary differential equations. 
As a result of numerical solution of this system, the conditions of the
  mutual compensation of the capillary and electrostatic forces on the conic
  surface of a liquid-metal anode have been obtained. 
This allows us to find the dependences of the cone angle and emission current 
  on the applied potential difference. 
They correctly represent the main features of the operation of a liquid-metal ion source. 
A comparison was made between the developed theoretical model and available
  experimental data for emission from pure gallium, indium, tin, or from alloys
  Au$+$Si, Co$+$Ge, and Au$+$Ge. 
Basing on our theoretical results, we have proposed explanations for some 
  specific features of the emissive behavior of these systems. 
So, the difference in behavior between indium and tin having close characteristics
  (the coefficient of surface tension, the mass/charge ratio) is probably related
  with the presence of doubly-charged ions Sn${}^{(2+)}$ for tin emission and
  with use of a sharper tungsten needle in the experiments with indium. 

In conclusion, let us note once again that the developed model does not pretend
  to describe all  aspects of operation of liquid-metal ion sources.
In particular, the consideration of the jet-like protrusion at the cone vertex,
  from where the vast majority of ions are evaporated,
  remains beyond the scope of the model.
The self-similar solutions used above describe LMIS operation in terms of
  the averaged characteristics ($\alpha$, $I$, and $U$) only.

\begin{acknowledgments}
The study was performed within the framework of the Program
  of Interdisciplinary Projects between the Ural Division and
  Siberian Branch of the Russian Academy of Sciences and the Program
  ``Basic Problems of Nonlinear Dynamics'' of the Presidium of the
  Russian Academy of Sciences.
It was financially supported by the Russian Foundation for Basic Research
  (project 07-02-96035), by the Presidential Program of Grants in Science
  (project MD-2553.2007.2), and by the Dynasty Foundation.
\end{acknowledgments}

\end {document}